\documentclass[aps,prd,nofootinbib,superscriptaddress,preprint,eqsecnum,showkeys,showpacs,preprintnumbers]{revtex4-2}
\usepackage{amsmath}
\usepackage{amssymb}
\usepackage{graphicx}
\usepackage{epsfig}
\usepackage{xcolor,color}
\usepackage{url}
\usepackage{bm}
\usepackage{mathrsfs}
\usepackage[utf8]{inputenc}
\usepackage{hyperref}
\usepackage{enumerate}
\usepackage{amsthm}
\usepackage{bbm}
\usepackage[normalem]{ulem}
\usepackage{upgreek}
\usepackage{tensor}
\usepackage{siunitx}
\usepackage{orcidlink}
\usepackage[caption=false]{subfig}
\usepackage{colortbl}
\usepackage{hyperref} 
\usepackage{appendix} 
\usepackage{cleveref}
\usepackage{multirow}
\usepackage{booktabs}
\usepackage{threeparttable}
\usepackage[export]{adjustbox}
\usepackage{times}
\usepackage{commath}
 
 \allowdisplaybreaks[3]
\hypersetup{colorlinks=true,linkcolor=blue,anchorcolor=blue,citecolor=blue}  
\definecolor{colour1}{HTML}{0571b0} 
\definecolor{colour2}{HTML}{92c5de} 
\definecolor{colour3}{HTML}{f4a582} 
\definecolor{colour4}{HTML}{ca0020} 
\definecolor{colour5}{HTML}{fe4a49} 
\definecolor{colour6}{HTML}{2d3092} 

\hypersetup{colorlinks=true, linkcolor=colour6, citecolor=colour6,
filecolor=colour6, urlcolor=colour6}

\theoremstyle{plain}

\newcommand{\bea}{\begin{eqnarray*}}
\newcommand{\eea}{\end{eqnarray*}}
\newcommand{\bean}{\begin{eqnarray}}
\newcommand{\eean}{\end{eqnarray}}

\newcommand{\be}{\begin{equation}}
\newcommand{\ee}{\end{equation}}

\newcommand\beq{\begin{equation}}
\newcommand\eeq{\end{equation}}
\def\bea{\begin{eqnarray}}
\def\eea{\end{eqnarray}}

\begin{document}

\title{Finite-nuclear-size effect for hydrogenlike ions under high external pressure}

\author{Dengshan~Liu}
 \affiliation{College of Physics, Jilin University, Changchun 130012, China}
 
\author{Huihui~Xie}
 \affiliation{College of Physics, Jilin University, Changchun 130012, China}

\author{Pengxiang~Du}
 \affiliation{College of Physics, Jilin University, Changchun 130012, China}

 \author{Tianshuai~Shang}
 \affiliation{College of Physics, Jilin University, Changchun 130012, China}

\author{Jian~Li \footnote{Corresponding author:: jianli@jlu.edu.cn}}
 \affiliation{College of Physics, Jilin University, Changchun 130012, China}
 \affiliation{Institute of Theoretical Physics, Chinese Academy of Sciences, Beijing 100190, China}

\author{Jiguang~Li}
 \affiliation{Institute of Applied Physics and Computational Mathematics, Beijing 100088, China}

\author{Tomoya~Naito}
 \affiliation{Department of Nuclear Engineering and Management, Graduate School of Engineering, The University of Tokyo, Tokyo 113-8656, Japan}
  \affiliation{Department of Physics, Graduate School of Science, The University of Tokyo, Tokyo 113-0033, Japan}
 \affiliation{RIKEN Center for Interdisciplinary Theoretical and Mathematical Sciences (iTHEMS), Wako 351-0198, Japan}
\begin{abstract}
The influence of pressure on finite-nuclear-size (FNS) corrections to atomic energy levels and electron-capture decay rate is investigated in confined hydrogenlike ions. The ions are modeled inside an impenetrable spherical cavity, with a Gaussian distribution used to represent the nuclear charge distribution. For each confinement radius—used to simulate external pressure—the energies and wave functions of the lowest-lying bound states are determined by numerically solving the Dirac equation via the kinetically balanced generalized pseudospectral (KBGPS) method. In contrast to unconfined ions, both the FNS corrections and electron-capture decay rates increase markedly under pressure and exhibit parallel trends with increasing confinement. Pressure also removes level degeneracies and alters the relative magnitudes of FNS corrections across different bound states. Moreover, the nuclear charge radius is found to significantly affect the pressure-enhanced electron-capture decay rate.
\end{abstract}


\maketitle

\newpage



\section{Introduction}
In state-of-the-art electronic structure calculations of atoms and molecules~\cite{andrae2000413,andrae2002nuclear}, it is essential to consider the finite-nuclear-size effect, which accounts for the fundamental discrepancy between the conventional point-nucleus approximation and the physical reality of spatially extended nuclear charge distributions. For highly charged ions with a large nuclear charge, corrections due to the FNS effect on energy levels can be up to hundreds of electron volts~\cite{VMShabaev_1993,Deck_2005,Babak1,Valuev0652502,Xie042807,Kuzmenko052804}. Moreover, FNS effects on the weak decay processes of elementary particles involving electrons, i.e., $\beta$ decay~\cite{Niţescu_2020,PhysRevC.109.025501} and electron-capture decay~\cite{RevModPhys.49.77,Sarriguren_2020}, have been considered for a long time.

Hydrogenlike ions possess the simplest electronic structure, making them ideal systems for theoretical studies of FNS effects, and the investigation of FNS effects in these systems holds critical implications for fundamental research~\cite{PR200163,beier.88.011603,Sturm2014467,oreshkina.96.030501,volotka.113.023002,PINdel_2019,Blaum_2021,Paul.126.173001,Sailer2022479}. For example, FNS corrections to the bound-electron energy levels and $g$ factors of hydrogenlike ions could serve as critical benchmarks for stringent tests of quantum electrodynamics (QED)~\cite{volotka.113.023002,PINdel_2019,Blaum_2021,Paul.126.173001,Sailer2022479}. Additionally, accurate calculations of transition probabilities in hydrogenlike ions necessitate the inclusion of FNS corrections~\cite{Parpia,Popov2017366,Sommer2020}. Until now, for free hydrogenlike ions, the study of the FNS effect has been very comprehensive. Studying the variation of these quantities for hydrogenlike ions under external pressure is of great significance.

In 1937, the confined hydrogen atom model~\cite{MICHELS1937981} was proposed to study the variations in the polarizability and kinetic energy of a hydrogen atom subjected to high external pressure. This model assumes that the atomic nucleus is fixed at the center of an impenetrable spherical cavity, while the electron is constrained to move within this enclosed region. By varying the size of the enclosed region, the model can mimic the compression of the electron cloud under different external pressures.
Since this system does not require consideration of a large number of hydrogenlike ions and only focuses on a single hydrogenlike ion, it has attracted widespread attention over the subsequent decades. 
Studies have extensively examined the energy levels~\cite{Ley351,Guimarães_2005,PhysRevA.87.012502}, transition probabilities~\cite{Gold1992}, and polarizability of confined hydrogen atoms and hydrogenlike ions~\cite{CecilLaughlin_2004,MONTGOMERY20121992,PhysRevA.91.032507}. In Ref.~\cite{Aquino2018399}, the Schrödinger equation for the confined hydrogen atom was solved to investigate the dependence of finite-nuclear-size corrections on external pressure, which reveals that these corrections to the energy levels increase significantly with increasing external pressure. Given the notable impact of external pressure on FNS corrections to hydrogen atomic energy levels, together with the non-negligible relativistic effects in heavier hydrogenlike ions, it is imperative to utilize the Dirac equation to explore the corresponding relativistic FNS effects in hydrogenlike ions under these extreme conditions.

Moreover, the interiors of stellar plasmas are subjected to extreme pressure, extreme conditions where hydrogenlike ions become highly abundant~\cite{TAKAHASHI1983578}. The decay properties of such ions differ significantly from those of neutral atoms~\cite{PhysRevLett.62.1025,PhysRevLett.74.499,PhysRevLett.99.262501,PhysRevC.77.014306,PhysRevC.78.025503,WINCKLER200936,PhysRevC.84.014301}, and understanding their decay modes is crucial for clarifying key processes in stellar nucleosynthesis. Experimental studies have been conducted on various hydrogenlike ions~\cite{WINCKLER200936,Bosch_2013,OZTURK2019134800}; for instance, $\beta ^{+}$ and electron-capture decays have been investigated in the hydrogenlike ion $\mathrm{^{140}Pr^{58+}}$~\cite{PhysRevLett.99.262501}. Although radioactive decay is generally considered to be independent of the external environment, the electron-capture decay rate $(\lambda _{\mathrm{EC} })$ is an exception. It depends strongly on external conditions~\cite{Hen1164,LIU2000163,LEE2008628,Ray2021,PRL2004,PRL2007,PRB2008,PRC2024}, since it is proportional to the electron density at the nucleus~\cite{RevModPhys.49.77}. Therefore, it is crucial to study the influence of external pressure on electron-capture decay in hydrogenlike ions.

In this paper, the effects of pressure on FNS corrections to energy levels and the decay rate of electron-capture decay are systematically investigated by means of the confinement model. The paper is organized as follows. In Sect.~\ref{sec:2}, the point-nucleus Coulomb potential and the Gaussian charge distribution model potential for confined hydrogenlike ions are introduced, along with the definition of the FNS correction for atomic energy levels under pressure. In Sect.~\ref{sec:3} and Sect.~\ref{sec:4}, the FNS corrections to the energy levels and the decay rate of electron-capture decay for hydrogenlike ions under high pressure are systematically analyzed, respectively. Finally, conclusions and an outlook are presented in Sect.~\ref{sec:5}.


\section{Theoretical method}\label{sec:2}

The effects of pressure on FNS corrections to energy levels and the decay rate of electron-capture decay are studied using the confinement model. The hydrogenlike ion under isotropic pressure is modeled by confining it within an impenetrable spherical box with confinement radius $R_{0}$. The corresponding point-nucleus Coulomb potential reads
\begin{equation}
\label{1}
     V(r)=\left\{\begin{matrix}-\frac{eZ}{r},  & r < {R_{0}},   \\ \infty,   & r \ge {R_{0}},\end{matrix}\right.
\end{equation}
where $Z$ is the atomic number.
To consider the finite nuclear charge density distributions, the Gaussian charge distribution model~\cite{Xie042807} is used. It should be noted that although more accurate nuclear charge densities are available~\cite{shangts2024,RN137,yundong2025}, a simple Gaussian charge density distribution model featuring an analytical Coulomb potential is adopted to facilitate discussions and calculations here. The corresponding potential with impenetrable walls is expressed as
\begin{equation}
\label{2}
 V(r)=\left\{\begin{matrix}-\frac{eZ}{r}\mathrm{erf} (\sqrt{\xi } r ),  & r < {R_{0}},   \\ \infty,  & r \ge {R_{0}}\end{matrix}\right.
\end{equation}
with $\xi =\frac{3}{2r_{c}^{2} }$, where $r_{c}$ is the root-mean-square nuclear charge radii and $\mathrm{erf} (x)$ is the error function.
For the relativistic Dirac equation of the one-electron system, i.e., hydrogenlike ion,  after separating out the angular parts of the wave function, the radial Dirac equation is given by:
\begin{equation}  
\label{2.3}
\begin{pmatrix}V(r)  & -c(\frac{d}{dr}-\frac{\kappa }{r}  )\\c(\frac{d}{dr}+\frac{\kappa }{r}  )  & V(r)-2c^{2}\end{pmatrix} \begin{pmatrix}P_{n\kappa}(r) \\Q_{n\kappa }(r)\end{pmatrix} =E \begin{pmatrix}P_{n\kappa}(r) \\Q_{n\kappa }(r)\end{pmatrix} .
\end{equation}
Substituting the above potential fields, Eq. \eqref{1} or Eq. \eqref{2}, into the radial Dirac equation \eqref{2.3} yields the energy eigenvalues $E_{\mathrm{Gauss}}[n\kappa]$ or $E_{\mathrm{point}}[n\kappa]$ ~\cite{Xie042807}, respectively, as well as the electron density at the nucleus, $\rho(0)$.
The FNS correction to atomic energy levels under pressure is  expressed as 
\begin{equation}
    \bigtriangleup E_{\mathrm{FNS}}=E_{\mathrm{Gauss}}-E_{\mathrm{point}}.
\end{equation}

Additionally, the hydrostatic pressure $P$ exerted on the electronic system by confinement effects can be estimated based on the variation of the ground-state energy with changes in the confinement volume~\cite{PhysRevA.87.012502}, i.e.,
\begin{equation}
\label{2.5}
    P=-\frac{1}{4\pi R_{0}^{2}}\frac{d E}{d R_{0}}. 
\end{equation}
The electron-capture decay rate $(\lambda _{\mathrm{EC} })$ is proportional to the electron density at the nucleus, i.e., $\lambda_{\mathrm{EC}} \propto  \rho (0)$, 
and the variation of $\rho(0)$ depends on the confinement radius $R_0$. The relationship between the external pressure and the confinement radius can be described by Eq. \eqref{2.5},
from which the relationship between the fractional increase of the decay rate~\cite{Hen1164,LIU2000163} and the external pressure can be derived~\cite{Ray2021}: 
\begin{equation}
\frac{\lambda -\lambda_{0} }{\lambda_{0} } =\frac{\rho (0)-\rho (0)_{\mathrm{ref}}}{\rho (0)_{\mathrm{ref}}} ,
\end{equation}
where $\lambda_{0}$ is the reference electron-capture decay rate at zero pressure, $\rho (0)_{\mathrm{ref}}$ is the corresponding electron density at the nucleus when there is no pressure, and $\lambda$ is the decay rate under pressure.
Both $\rho(0)_{\mathrm{ref}}$ and $\rho(0)$ are derived from solving the Dirac
equation; therefore, they inherently account for relativistic effects.

\section{Pressure effects on FNS corrections to energy levels}\label{sec:3}

\begin{figure}[b]
\centering
\includegraphics[width=0.45\textwidth]{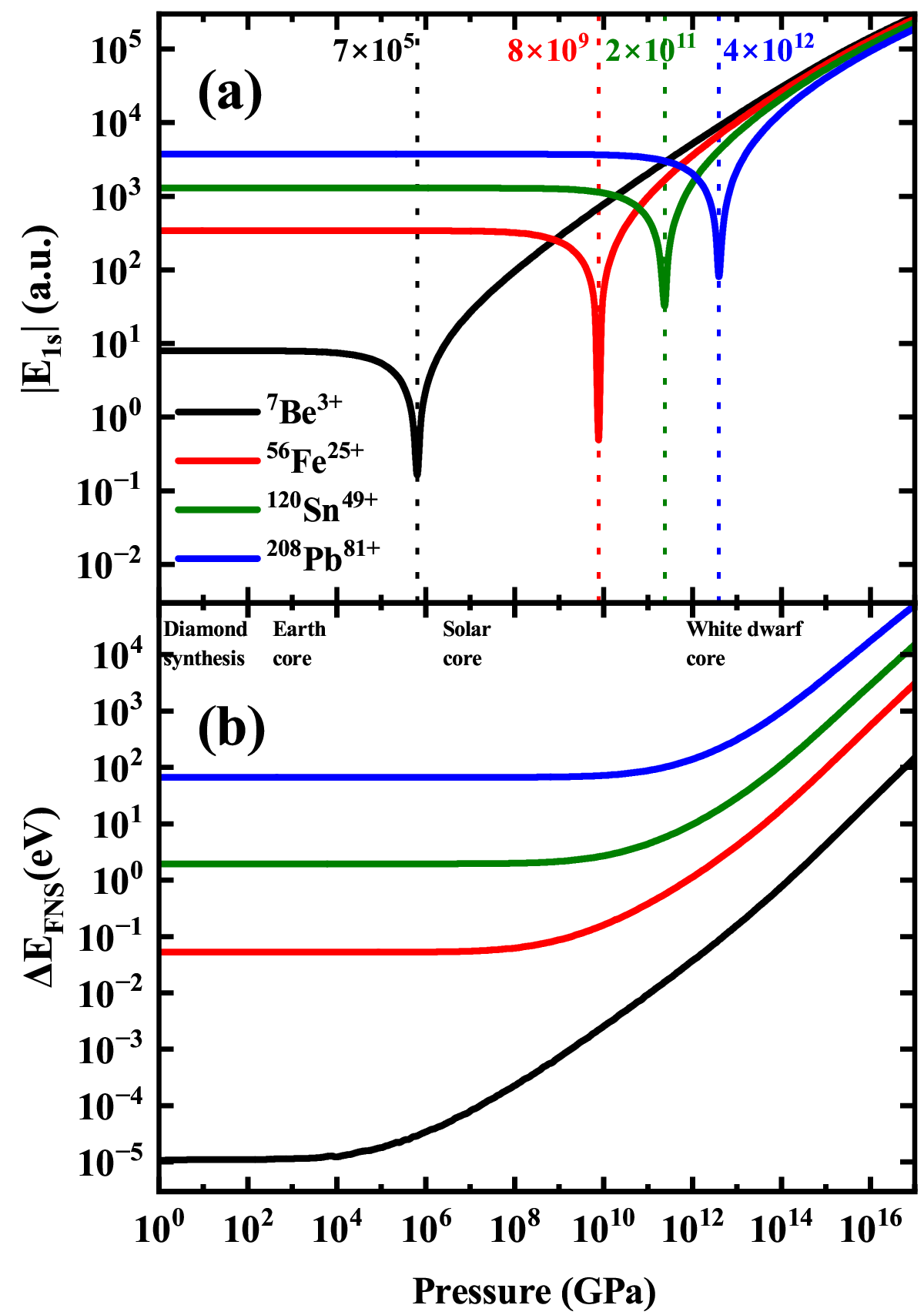}
\caption
{The absolute energy values $|E_{1s}|$ (a) and the corresponding FNS correction $\Delta E_{\mathrm{FNS} } $ (b) for the $1s_{1/2}$ state of hydrogenlike ions $^{7}\mathrm{Be}^{3+}$, $^{56}\mathrm{Fe}^{25+}$, $^{120}\mathrm{Sn}^{49+}$ and $^{208}\mathrm{Pb}^{81+}$ as a function of pressure. The vertical dashed line of (a) represents the pressure where the energy level is zero. At the top of (b), some characteristic pressures are labeled that represent several orders of magnitude.}
\label{fig:figure1}
\end{figure}

Figure \ref{fig:figure1} (a) and (b) display the absolute energy values $|E_{1s}|$ — which incorporates the FNS effect — and corresponding FNS correction to energy levels $\Delta E_{\mathrm{FNS} }$ for the ground state of hydrogenlike ions $^{7}\mathrm{Be}^{3+}$, $^{56}\mathrm{Fe}^{25+}$, $^{120}\mathrm{Sn}^{49+}$ and $^{208}\mathrm{Pb}^{81+}$, respectively, as a function of pressure. The calculations cover a pressure range from 1 GPa to $10^{17}$ GPa. Owing to the significant variations in both the $|E_{1s}|$ and $\Delta E_{\mathrm{FNS} }$ across this pressure range, a logarithmic scale is used for the vertical axis in Fig. \ref{fig:figure1} to better represent their behavior, and the absolute values of the energy levels are plotted.

Figure \ref{fig:figure1}(a) clearly illustrates that the ground-state energy $E_{1s}$ increases monotonically with pressure. 
At low pressures, the energy levels approach those of free hydrogenlike ions asymptotically, remaining negative. As pressure increases, the energies rise gradually toward zero. With further increase in pressure, the energy values eventually exceed zero and exhibit a rapid increase.
This behavior arises because higher pressure leads to greater localization of the corresponding wave function. According to the Heisenberg uncertainty principle, enhanced localization increases momentum.
The pressure at which the energy level reaches zero is defined as the critical pressure, indicated by the dashed lines in Fig. \ref{fig:figure1}(a). On both sides of this critical point, the energy levels display markedly different trends. A comparison of critical pressures among different hydrogenlike ions reveals that those with smaller nuclear charges have lower critical pressures. For instance, the critical pressure increases from $7\times 10^{5}$ $\mathrm{GPa} $ in $^7\mathrm{Be} ^{3+}$ to $8\times 10^{9}$ $\mathrm{GPa} $ in $^{56}\mathrm{Fe}^{25+}$, $2\times 10^{11}$ $\mathrm{GPa} $ in $^{120}\mathrm{Sn}^{49+}$ and $4\times 10^{12}$ $\mathrm{GPa} $ in $^{208}\mathrm{Pb}^{81+}$. This is because hydrogenlike ions with smaller nuclear charges possess ground state wavefunctions distributed over a larger spatial extent, rendering them more susceptible to compression under pressure and thus more sensitive to external confinement.


It can be observed from Fig. \ref{fig:figure1}(b) that when the pressure is below the critical pressure, $\Delta E_{\mathrm{FNS} }$ increases very slowly. As the pressure approaches the critical pressure, $\Delta E_{\mathrm{FNS} }$ starts to increase rapidly and continues to grow with further pressure increases. This indicates that as the pressure increases, the FNS effect becomes progressively more significant.
Although the ground state of hydrogenlike ions is bound by an infinite potential barrier rather than the Coulomb potential when the pressure exceeds the critical value, the strong external pressure compresses the electron wavefunction toward the nucleus, thereby enhancing the FNS effect.
Lighter hydrogenlike ions are more sensitive to pressure; the magnitude of $\Delta E_{\mathrm{FNS} }$ in light hydrogenlike ions increases much faster than in heavy hydrogenlike ions. For example, $\Delta E_{\mathrm{FNS} }$ in $ ^{7}\mathrm{Be}^{3+}$ is about seven orders of magnitude smaller than in $^{208}\mathrm{Pb}^{81+}$ at approximately 100 GPa, and it finally becomes three orders of magnitude smaller as pressure approaches $10^{16}$ GPa. In Fig. \ref{fig:figure1}(b), the orders of magnitude of several characteristic pressures are presented to give an idea of the pressures involved.

\begin{figure}
\centering
\includegraphics[width=0.45\textwidth]{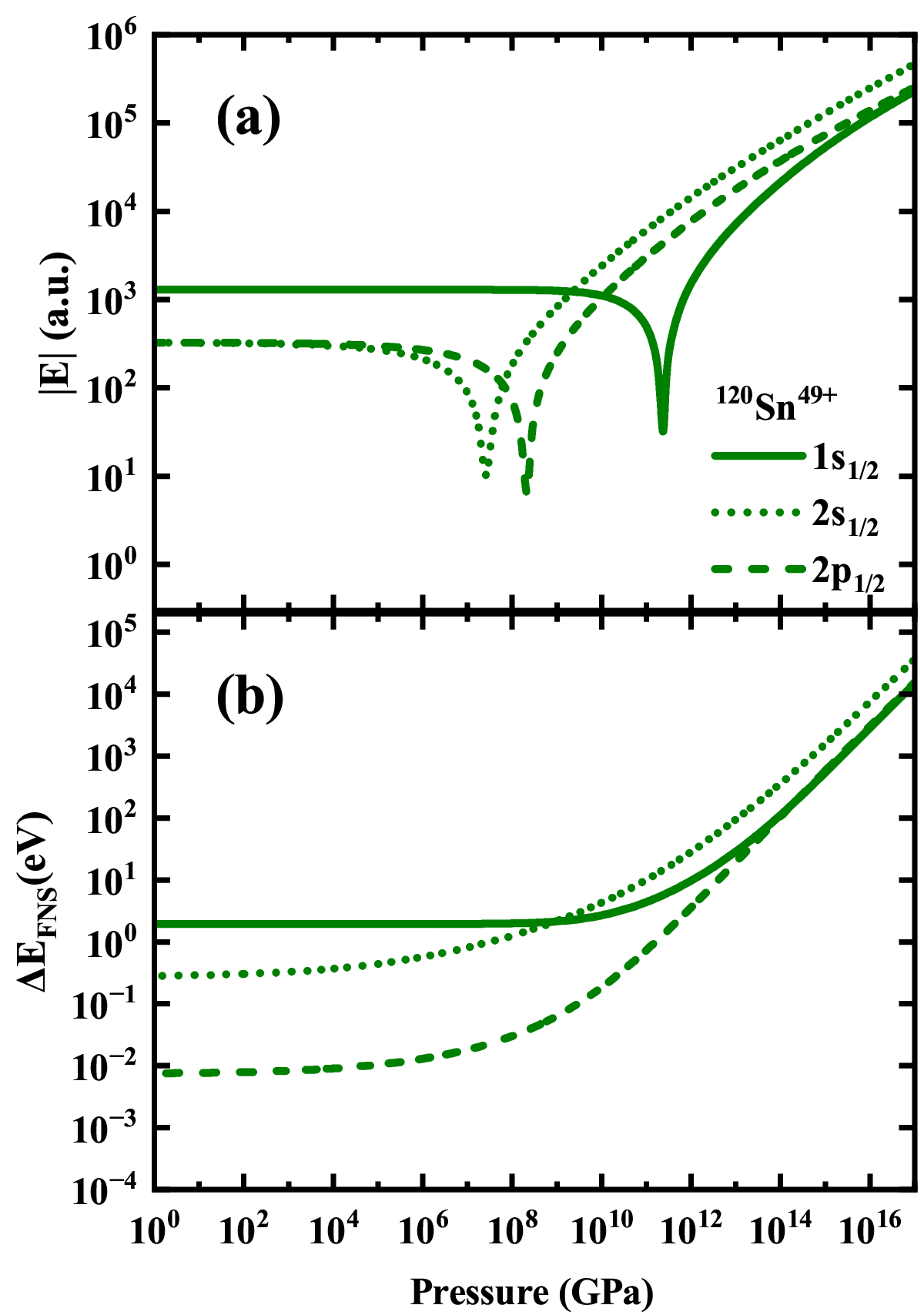}
\caption{Absolute values of energy levels (a) and FNS corrections to energy levels $\Delta E_{\mathrm{FNS} } $  (b) for $ 1s_{1/2}$, $2s_{1/2}$ and $2p_{1/2}$ states of hydrogenlike ion $ ^{120}\mathrm{Sn}^{49+}$ as a function of pressure.}
\label{fig:figure2}
\end{figure}

It is meaningful to investigate the effect of pressure on different bound states. Fig. \ref{fig:figure2}(a) and (b) present the absolute values of energy levels and corresponding FNS corrections $\Delta E_{\mathrm{FNS} } $ as a function of pressure for $ 1s_{1/2}$, $2s_{1/2}$ and $2p_{1/2}$ states in hydrogenlike ion $ ^{120}\mathrm{Sn}^{49+}$. It can be observed from Fig. \ref{fig:figure2} that the critical pressures of excited states of the hydrogenlike ion $ ^{120}\mathrm{Sn}^{49+}$ are lower than those of the ground state. This is attributed to the fact that the energy levels of the excited states are higher than those of the ground state, and with increasing pressure, the energy levels of the excited states converge toward zero more swiftly. Furthermore, the critical pressures of the $2s_{1/2}$ and $2p_{1/2}$ states are relatively close; when the pressure approaches their critical pressures, about $10^6$ GPa, the difference between their energy levels becomes increasingly prominent. Subsequently, with further increases in pressure, the energy level of $2s_{1/2}$ state remains higher than that of the $2p_{1/2}$ state. However, once the pressure exceeds the critical pressure of the $1s_{1/2}$ state, the energy level of $1s_{1/2}$ state gradually approaches that of the $2p_{1/2}$ state, driven by increasingly prominent relativistic effects under high-pressure conditions.

In Fig. \ref{fig:figure2}(b), by comparing FNS corrections $\Delta E_{\mathrm{FNS} } $ for different bound states, it can be observed that at low pressure, the $\Delta E_{\mathrm{FNS} } $ for $ 1s_{1/2}$ state is higher than that for $ 2s_{1/2}$ state, and the $\Delta E_{\mathrm{FNS} } $ for $ 2s_{1/2}$ state is higher than that for $ 2p_{1/2}$ state. This is consistent with the statement in Ref.~\cite{Xie042807}. However, when the pressure reaches $10^{9}$ GPa, the $\Delta E_{\mathrm{FNS} } $ in $2s_{1/2}$ state exceeds that in the $1s_{1/2}$ state, and as the pressure increases further, the difference in $\Delta E_{\mathrm{FNS} } $ between the two states continues to grow. Meanwhile, $\Delta E_{\mathrm{FNS} } $ in the $2p_{1/2}$ state gradually approaches that in the $1s_{1/2}$ state, and after the pressure exceeds $10^{14}$ GPa, $\Delta E_{\mathrm{FNS} } $ of the two states become nearly equal. Therefore, from this figure, it can be seen that the relative magnitudes of $\Delta E_{\mathrm{FNS} } $ for different bound states undergo significant changes with increasing pressure.

\begin{figure}
\centering
\includegraphics[width=0.45\textwidth]{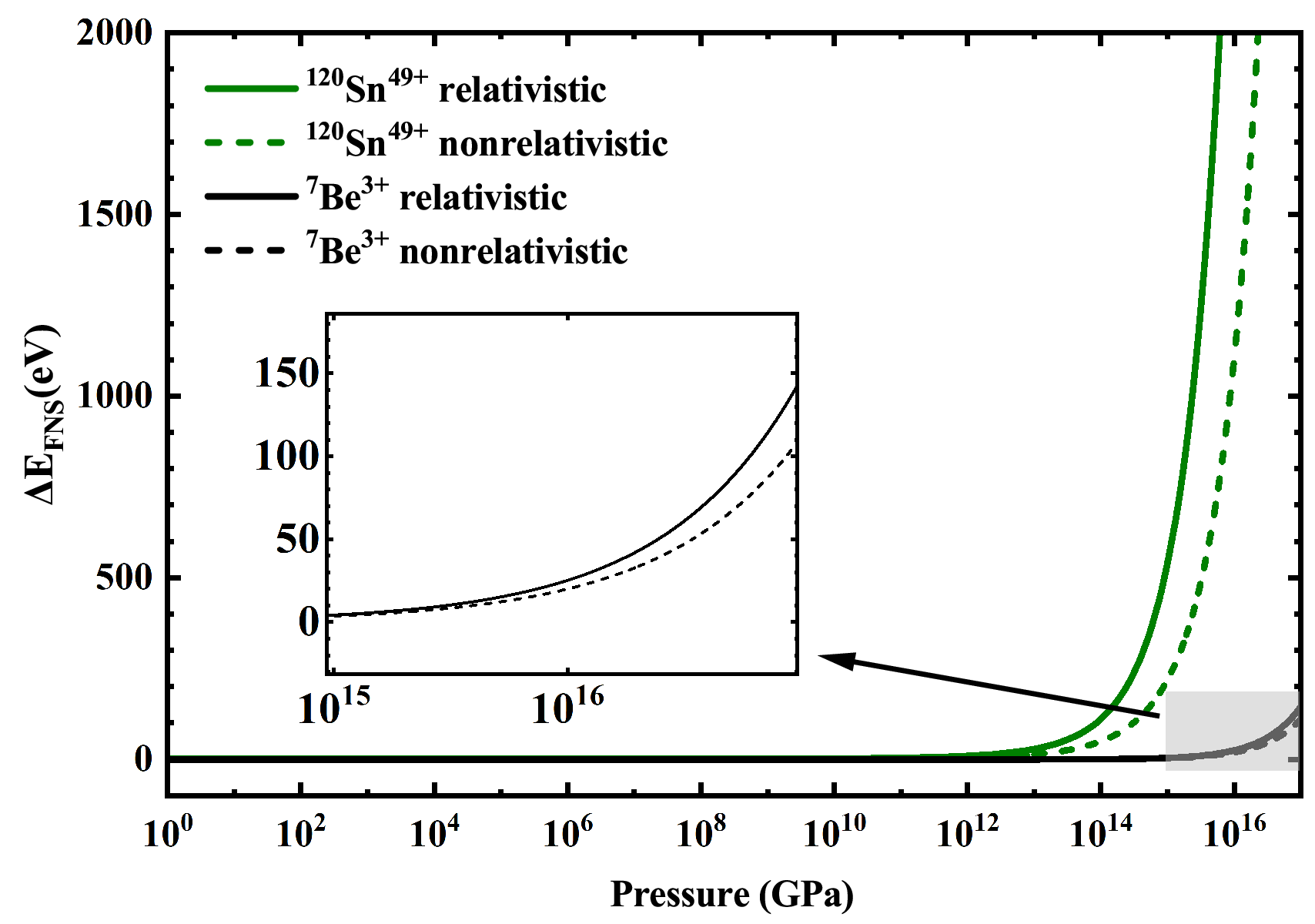}
\caption{$\Delta E_{\mathrm{FNS} } $ for the $1s_{1/2}$ state of hydrogenlike ions $^{7}\mathrm{Be}^{3+}$ and $^{120}\mathrm{Sn}^{49+}$ in the relativistic case, are compared with the nonrelativistic case.}
\label{fig:figure3}
\end{figure}

In studies of confined atomic systems, the nonrelativistic Schrödinger equation is commonly used. However, this formulation does not account for relativistic effects, which play a significant role in hydrogenlike ions with high nuclear charge or under high-pressure conditions. It is therefore essential to evaluate the influence of relativistic corrections on $\Delta E_{\mathrm{FNS}}$. To address this, the nonrelativistic Schrödinger equation for confined hydrogen-like ions is solved considering both Gaussian and point-like nuclear charge distributions, yielding the corresponding nonrelativistic values of $\Delta E_{\mathrm{FNS}}$. These results are subsequently compared with those derived from the Dirac equation. The numerical solution of the Schrödinger equation is obtained via the generalized pseudospectral (GPS) method. 
The GPS method is a discrete variable representation approach based on global basis functions and Gauss-type quadratures~\cite{Jiao022801}, which transforms the differential equation into a matrix eigenvalue problem by mapping the infinite domain onto a finite interval.
Implemented within the discrete-variable representation framework, the GPS method demonstrates rapid convergence and high versatility in solving quantum eigenvalue problems~\cite{CHU20041,Xie26653,Jiao_2021,ZhuLin2020,Jiao022801}. Furthermore, significant progress has also been made in solving the Dirac and Schrödinger equations using machine learning~\cite{Wang2025,wang2024}.


Figure \ref{fig:figure3} compares the FNS corrections to energy levels $\Delta E_{\rm FNS}$ for the $1s_{1/2}$ state of hydrogenlike ions $^{7}\mathrm{Be}^{3+}$ and $^{120}\mathrm{Sn}^{49+}$ between the relativistic and nonrelativistic cases. 
The relativistic $\Delta E_{\rm FNS}$ is consistently greater than its nonrelativistic counterpart. Furthermore, the disparity in $\Delta E_{\rm FNS}$ attributable to relativistic effects intensifies significantly with increasing pressure. 
Beyond pressures of $10^{14}$ GPa, this difference becomes particularly pronounced for the hydrogenlike ion $^{120}\mathrm{Sn}^{49+}$. In contrast, for $^{7}\mathrm{Be}^{3+}$ within the same pressure range, the differences are considerably smaller than those observed for $^{120}\mathrm{Sn}^{49+}$. Consequently, accounting for relativistic effects is essential for hydrogenlike ions with higher nuclear charges under extreme pressure conditions.

\begin{figure}
\centering
\includegraphics[width=0.45\textwidth]{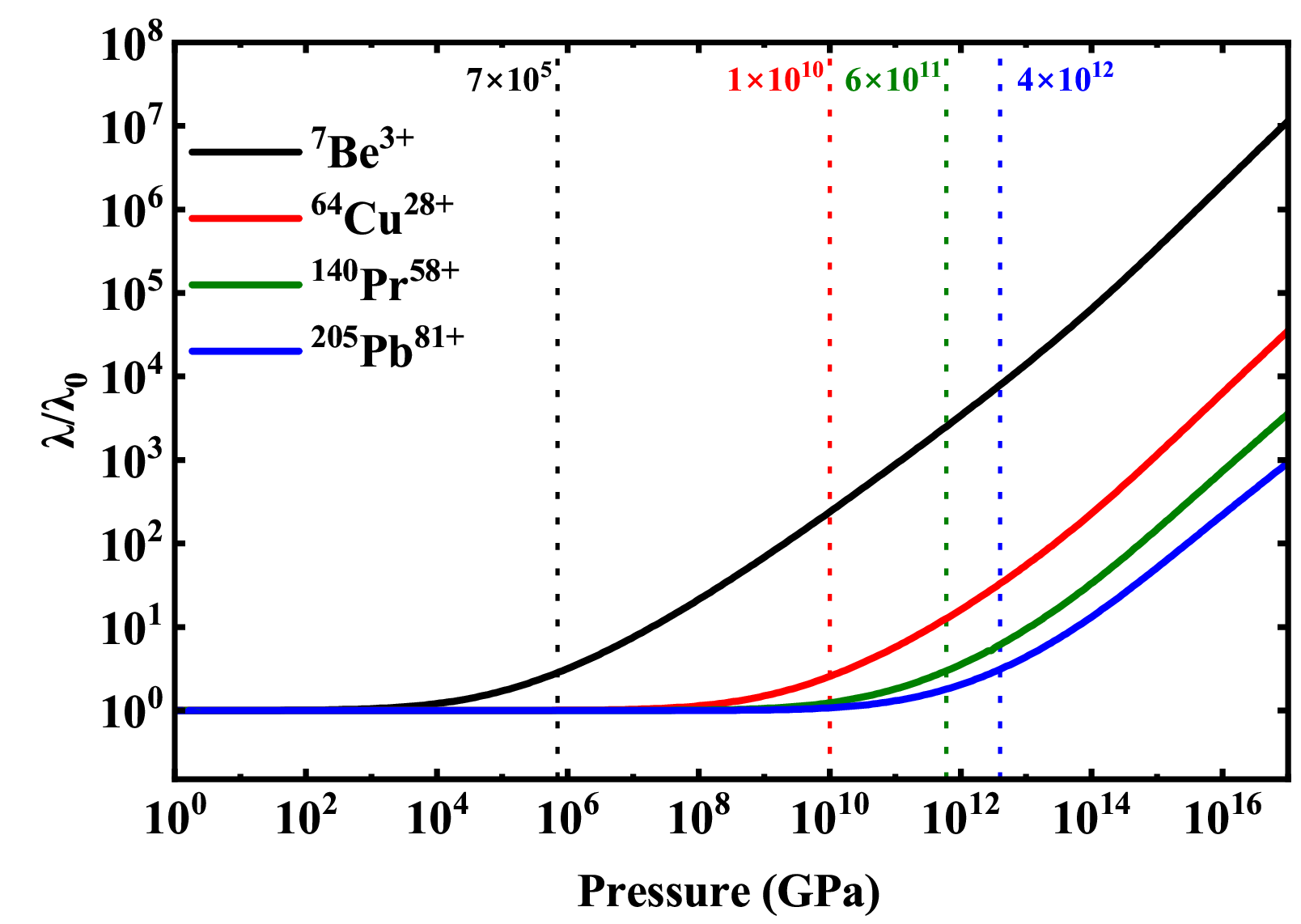}
\caption{The ratio of the electron-capture decay rate with and without pressure for the $1s_{1/2}$ state of hydrogenlike ions $ ^{7}\mathrm{Be}^{3+}$, $^{64}\mathrm{Cu}^{28+}$, $^{140}\mathrm{Pr}^{58+}$ and $^{205}\mathrm{Pb}^{81+}$ as a function of pressure. The vertical dashed line denotes the critical pressure for the corresponding hydrogenlike ions.}
\label{fig:figure4}
\end{figure}

\section{The effect of pressure on electron-capture decay}\label{sec:4}


Figure \ref{fig:figure4} displays the ratio between the electron-capture decay rate  of $ ^{7}\mathrm{Be}^{3+}$, $^{64}\mathrm{Cu}^{28+}$, $^{140}\mathrm{Pr}^{58+}$ and $^{205}\mathrm{Pb}^{81+}$ under pressure and that at zero pressure. By observing the relationship between the ratio and pressure, it is evident that the variation of the electron-capture decay rate under pressure exhibits a significant similarity to the changes in the FNS corrections to energy levels $\Delta E_{\rm FNS}$ under pressure. For example, when the pressure is below the critical pressure, there is no significant change in the electron-capture decay rate. However, as the pressure approaches the critical pressure, the electron-capture decay rate increases rapidly with the pressure. The ratio of the electron-capture decay rate for hydrogenlike ion $ ^{7}\mathrm{Be}^{3+}$ has increased by approximately seven orders of magnitude within the current pressure range. 
Moreover, as the pressure reaches to $10^6$ GPa (approximately equal to the pressure at the solar core), the electron-capture decay rate of $^{7}\mathrm{Be}^{3+}$ ions has already undergone a significant change; thus, this is likely to exert an important influence on the hydrogen burning process inside the Sun. For nuclides with a larger proton number, a significant change in the electron-capture rate would only be possible at higher pressures.

\begin{figure}
\begin{center}
\includegraphics[width=0.9\linewidth,angle=0]{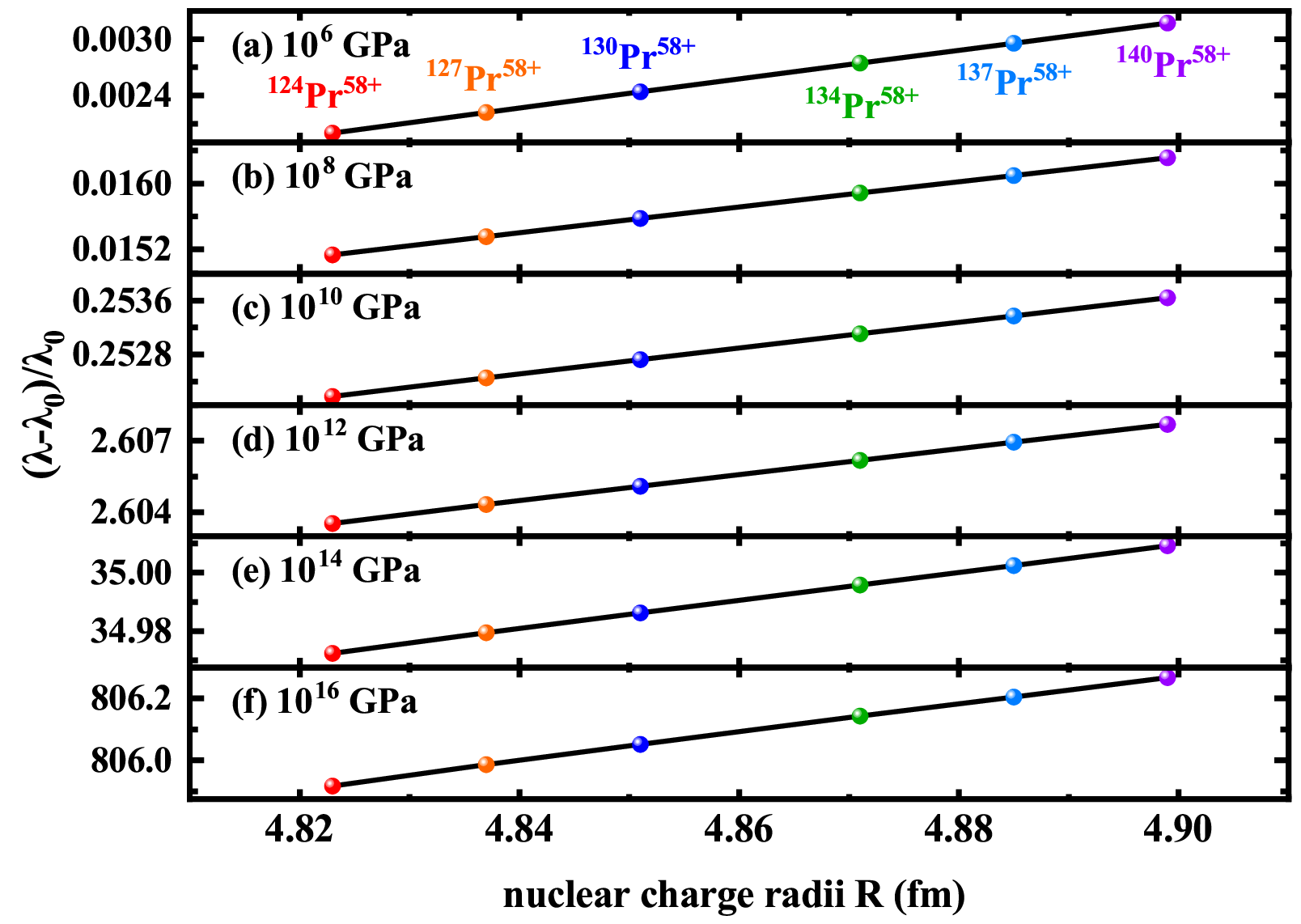}
\caption{The fractional increase of the decay rate at the $1s_{1/2}$ state of hydrogenlike ions of Pr isotopes. (a) Fixed different pressures at $10^6$ GPa, (b) $10^8$ GPa, (c) $10^{10}$ GPa, (d) $10^{12}$ GPa, (e) $10^{14}$ GPa, and (f) $10^{16}$ GPa.}
\label{fig:figure5}
\end{center}
\end{figure}

As the electron density at the nucleus $\rho (0)$ depends not only on external pressure but also critically on the nuclear charge radius, the latter plays a significant role in modulating $\rho (0)$. The variation in nuclear charge radii across an isotope chain motivates a detailed investigation of the electron-capture decay rates among these nuclides under high external pressure. Figure \ref{fig:figure5} depicts $(\lambda -\lambda _{0} )/\lambda _{0}$ for the $1s_{1/2}$ state of hydrogenlike ions of Pr isotopes at several different pressure magnitudes. It can be observed that $(\lambda -\lambda _{0} )/\lambda _{0}$ rises with the increase of nuclear charge radius. This indicates that pressure has a more significant effect on isotopes with larger nuclear charge radii.

In Fig. \ref{fig:figure5}(a), the pressure is $10^{6}$ GPa, which is roughly equivalent to the pressure at the core of the sun, $(\lambda -\lambda _{0} )/\lambda _{0}$ increases from 0.002 to 0.003 as the nuclear charge radii $r_c$ increases from 4.823 to 4.899 fm. This indicates that, within the current pressure range, the influence of the nuclear charge radius on the decay rate is roughly comparable to the effect of external pressure on that. However, as the pressure increases, the influence of pressure on the decay rate gradually becomes dominant. 

\section{ Conclusions and discussions}\label{sec:5}
In summary, the pressure dependence of FNS effects and electron-capture decay rates has been investigated for different hydrogenlike ions by using the confined atom model. It is found that both the FNS energy corrections $\Delta E_{\rm FNS}$ and the electron-capture decay rate increase monotonically as pressure increases. Based on the pressure-induced evolution of energy levels, a critical pressure was defined, which is shown to play a key role in governing the behavior of both FNS corrections and decay rates under compression. 
The magnitudes of FNS corrections to energy levels for $1s_{1/2}$, $2s_{1/2}$ and $2p_{1/2}$ states change as pressure increases. 
Additionally, it is essential to consider the relativistic effects for the $\Delta E_{\rm FNS}$ under extreme pressure conditions. Investigation of Pr isotopes revealed that isotopes with larger nuclear charge radii exhibit a greater fractional increase in electron-capture decay rate under pressure.


Our results show that certain nuclear-related quantities undergo significant modifications under high pressure. 
Existing experimental studies have already demonstrated that confinement effects can alter the electron-capture decay rate of beryllium~\cite{PRL2004,PRL2007,PRB2008,PRC2024}. Therefore, future studies will extend this methodology by employing the confined atom model to investigate the effects of pressure on electron-capture decay in multi-electron systems. Furthermore, research on high-pressure single-atom systems holds promise as a prospective research direction in the field of high-pressure physics. Under high-pressure conditions, numerous atoms form solid phases, a subset of which are metallic~\cite{Wigner1935}. Accounting for such phase transitions in the context of electron-capture decay rates is also essential, and this aspect is reserved for future investigations.

\vspace{0.3cm}
\acknowledgments
{ Dengshan Liu, Huihui Xie, Pengxiang Du, Tianshuai Shang and Jian Li acknowledge the National Natural Science Foundation of China  (Grants No. 12475119 and 12447101) and the China Postdoctoral Science Foundation, No. 2025M783390.
Tomoya Naito acknowledges the JSPS Grant-in-Aid under Grant Nos. JP23H01845, JP23K03426, JP23K26538, JP24K17057, JP25H00402, JP25H0158, JP25K01003, JP25KJ0405 and JST COI-NEXT Grant No. JPMJPF2221.}  


\bibliography{bibliography}

\end{document}